\documentclass[a4paper]{jpconf}
\usepackage{graphicx}
\usepackage{amssymb,amsthm}

\def\one{\mbox{1 \kern-.59em {\rm l}}}

\newcommand{\R}{{\mathbb{R}}}
\newcommand{\N}{{\mathbb{N}}}
\newcommand{\C}{{\mathbb{C}}}
\newcommand{\Z}{{\mathbb{Z}}}

\def\mg{\mathfrak{g}}
\def\mh{\mathfrak{h}}
\def\mr{\mathfrak{r}}

\def\ms{\mathfrak{s}}
\def\mun{\mathfrak{u}}

\def\diag{\mbox{diag}}

\def\ii{{\,{\rm i}\,}}
\def\ddd{{\rm d}}
\def\eee{{\,\rm e}\,}
\def\vol{{\rm vol}}
\def\pfaff{\mbox{pfaff}}

\def\cN{{\cal N}}
\def\cO{{\cal O}}
\def\cC{{\cal C}}
\def\cJ{{\cal J}}
\def\cW{{\cal W}}

\def\nn{\nonumber}
\def\bea{\begin{eqnarray}}
\def\eea{\end{eqnarray}}
\def\be{\begin{equation}}
\def\ee{\end{equation}}

\newcommand{\eq}[1]{(\ref{#1})}

\def\a{\alpha}
\def\b{\beta}

\newtheorem{theorem}{Theorem}

\begin{document}

\begin{flushright}
UWThPh--2007--18\\
HWM--07--28\\
EMPG--07--17\\
\end{flushright}

\title{Nonabelian localization for \\ gauge theory on the fuzzy
  sphere\footnote{Based on talk given by H.S. at the International
  Conference ``Noncommutative Geometry and Physics'', April~23--27,
  2007, Orsay, France. To be published in {\sl Journal of Physics
    Conference Series}.}}

\author{Harold Steinacker$^1$, Richard J. Szabo$^2$}

\address{$^1$ Fakult\"at f\"ur Physik,
Universit\"at Wien \\ Boltzmanngasse 5, A-1090 Wien, Austria}
\address{$^2$ Department of Mathematics and \\ Maxwell
  Institute for Mathematical Sciences \\ Heriot-Watt University,
Colin Maclaurin Building \\ Riccarton, Edinburgh EH14 4AS, U.K.}

\ead{harold.steinacker@univie.ac.at , R.J.Szabo@ma.hw.ac.uk}

\begin{abstract}
We apply nonabelian equivariant localization techniques to Yang-Mills
theory on the fuzzy sphere to write the partition function entirely as
a sum over local contributions from critical points of the action. The
contributions of the classical saddle-points are evaluated explicitly, 
and the partition function of ordinary Yang-Mills theory on the sphere
is recovered in the commutative limit.
\end{abstract}

\setcounter{equation}{0}\section{Introduction}

The formulation of field theories on noncommutative spaces is expected
to incorporate to some extent the effects of quantum gravity in a
field theoretic framework (see e.g. ~\cite{douglas,szabo} for 
reviews, and \cite{ncgrav,Szabo:2006wx} concerning the relation with gravity). 
Their quantization, however, is rather non-trivial, due to a
new phenomenon called UV/IR mixing. This problem appears to be very
generic in noncommutative field theories, both for scalar and for
gauge field theories. In essence it means that the ultraviolet
divergences not only lead to the usual infinite renormalizations of
the masses and couplings, but also to new divergences in the infrared
behaviour of propagators, which are likely to
signal new physics. It is therefore important to develop appropriate
techniques for the quantization of noncommutative field theories,
and to find models which are well-defined in order to avoid problems
which are possibly associated to mathematical artifacts.

Fuzzy spaces provide a nice class of noncommutative spaces based on
finite-dimensional algebras of ``functions'', with the same symmetries
as their classical counterparts. This means that field theory on fuzzy
spaces is naturally regularized, but the regularization is compatible
with a geometrical symmetry group (in contrast to lattice field
theory, for example). A large family of such spaces is given by the
quantization of coadjoint orbits $\cO$ of a Lie group in terms of
certain finite matrix algebras $\cO_N$. They are labelled by a 
noncommutativity parameter $\frac 1N$, and the classical space
is recovered in the large~$N$ limit. The simplest example is the fuzzy
sphere $S^2_N$, which has been studied in great detail; see
e.g.~\cite{madore}--\cite{matrixsphere} and references therein. There
are also extensively studied four-dimensional examples, such as
$S^2_N \times S^2_N$ and $\C P^2_N$~\cite{stroh}--\cite{S2S2paper}.

In this article, we review the application of nonabelian localization
techniques to Yang-Mills theory on $S^2_N$ developed
in~\cite{Steinacker:2007iq}. This provides, along
with~\cite{szabo-paniak}, one of the few examples where noncommutative
gauge theory can be solved exactly. We will explicitly evaluate the
partition function and show that it reduces to the expected one on
the classical sphere $S^2$ in the limit~$N\to\infty$.

\setcounter{equation}{0}
\section{Equivariant Localization and the Duistermaat-Heckman
  theorem\label{DHThm}}

Let $X$ be a compact $2n$-dimensional symplectic manifold with
symplectic two-form $\omega$. Assume that the circle group $U(1)$ acts
globally on $X$ via symplectomorphisms, generated by a Hamiltonian
vector field $V$ with
\be  
\ddd H=-\iota_V\omega=-\omega(V,-)
\ee
for some real-valued function $H$ on $X$. The Duistermaat-Heckman
theorem (see e.g.~\cite{szaboloc} and references therein) then
states that the classical partition function
\be 
Z = \int_X\,{\omega^n\over n!}~\eee^{-\beta \,H}
\ee
is given {\em exactly} by the semi-classical approximation, i.e. by
summing over all critical points $P_i$ of $H$:
\be
Z = \sum_i \, \frac{\eee^{-\beta\,H(P_i)}}{\a_i} \ .
\ee
Here $\a_i$ is the product of the weights of the representation of the
$U(1)$ action in the tangent space at $P_i$, which is formally given by
the equivariant Euler class $e_V(P_i)={\rm pfaff}\,\ddd V(P_i)$ of the
normal bundle to the critical point set in $X$. As such, it is the
fluctuation determinant determined by integration over an
infinitesimal neighbourhood of $P_i$. 

The subject of this paper is the application of a generalization of
this theorem to compute the partition function of Yang-Mills theory
on the fuzzy sphere. However, there are several complications which
require a more sophisticated version of the localization formula. 
First, the global symmetry group $U(1)$ is replaced by the gauge
group, which is nonabelian and usually infinite-dimensional; in the
fuzzy case it becomes a finite-dimensional unitary group. 
Second, the saddle-points are replaced by critical surfaces.
These complications can be handled using techniques from equivariant
cohomology, following the method in~\cite{witten} developed for
ordinary two-dimensional Yang-Mills theory. In fact, the
formal treatment in~\cite{witten} is realized in our setting in a 
rigorous, finite-dimensional framework. We will also take advantage of
some more recent techniques in~\cite{beasley-witten} which allow for the
explicit evaluation of the contributions from the classical solutions
of the Yang-Mills equations of motion.

\setcounter{equation}{0}
\section{The fuzzy sphere}

The fuzzy sphere $S^2_N$ \cite{madore} is a matrix approximation of the
usual sphere $S^2$. The algebra of functions on $S^2$, 
spanned by the spherical harmonics, is truncated at a given frequency.  
The algebra then becomes the finite-dimensional algebra of $N\times N$
matrices. More precisely,
let $N\in\N$, and let $\xi_i$, $i=1,2,3$ be the $N\times N$ hermitian
coordinate generators of the fuzzy sphere $S^2_N\cong{\rm Mat}_N$
which satisfy the relations
\be \label{FS-l}
\epsilon^{ij}{}_{k}\, \xi_i \,\xi_j = \ii \xi_k \qquad
 \mbox{and}\qquad \xi_i\, \xi^i  = \mbox{$\frac14$}\,\left(N^2-1
\right)~\one_N
\ee
where throughout repeated upper and lower indices are implicitly
summed over. The deformation parameter is $\frac1N$ and $S_N^2$
becomes the algebra of functions on the classical unit sphere $S^2$
in the limit $N\to\infty$.
The quantum space $S_N^2$ preserves the
classical invariance under global rotations as follows.     
The $\xi_i$ generate an $N$-dimensional representation of
the global $SU(2)$ isometry group. Under the adjoint action of
$SU(2)$, this representation decomposes covariantly into
$p$-dimensional irreducible representations $(p)$ of $SU(2)$ as
\be
{\rm Mat}_N\cong(1)\oplus(3)\oplus\cdots\oplus(2N-1) \ ,
\label{MatNSU2decomp}\ee
which are interpreted as fuzzy spherical harmonics.
This decomposition defines a natural 
map from $S^2_N$ to the space of functions 
on the commutative sphere.
The integral of a function
$f\in S_N^2$ over the fuzzy sphere is given by the trace of $f$,
which  coincides with the usual integral on $S^2$
\be
\Tr(f) = \frac N{4\pi}\,\int_{S^2}\,\ddd\Omega~f \ 
\label{fuzzyint}\ee
where the above map is understood.
Rotational invariance of the integral 
then corresponds to invariance of the matrix trace under the
adjoint action of $SU(2)$.

Following~\cite{matrixsphere}, let us combine the generators $\xi_i$
into a larger hermitian $\cN\times\cN$ matrix
\be
\Xi = \mbox{$\frac 12$}\, \one_N\otimes\sigma^0 + \xi_i \otimes\sigma^i
\label{Xi-collective}
\ee 
where $\cN =  2N$, $\sigma^0 = \one_2$, while
\be
\sigma^1=\left(\begin{array}{cc}0&1\\1&0\end{array}\right) \ , 
\quad\sigma^2=\left(\begin{array}{cc}0&\ii\\-\ii&0\end{array}\right)
 \quad \mbox{and}
\quad \sigma^3=\left(\begin{array}{cc}1&0\\0&-1\end{array}\right)
\label{Pauli}
\ee
are the Pauli spin matrices obeying
\be
\Tr\big(\sigma^i\big)=0 \qquad \mbox{and} \qquad
\sigma^i\,\sigma^j=\delta^{ij}\,\sigma^0+\ii\epsilon^{ij}{}_k\,\sigma^k
\ .
\label{Pauliids}
\ee
One easily finds from (\ref{FS-l}) and (\ref{Pauliids}) the identities
\be
\Xi^2 = \mbox{$\frac{N^2}4$}~\one_{\cN} \qquad \mbox{and} \qquad
\Tr(\Xi) = N \ .
\ee
Since $\xi_i\otimes\sigma^i$ is an intertwiner of the Clebsch-Gordan
decomposition $(N)\otimes(2)=(N-1)\oplus(N+1)$, this implies that
$\Xi$ has eigenvalues $\pm \,\frac N2$ with respective multiplicities
$N_\pm=N\pm1$.

\setcounter{equation}{0}
\section{Gauge theory on the fuzzy sphere}
\label{sec:gauge}

\subsection{Configuration space}

We will now describe the gauge field degrees of freedom in our
formulation. To elucidate the construction in as transparent a way as
possible, we begin with the abelian case of $U(1)$ gauge theory. To
introduce $\mun(1)$ gauge fields $A_i$ on $S_N^2$, consider the
covariant coordinates
\be
C_i = \xi_i + A_i \qquad \mbox{and} \qquad C_0 = 
\mbox{$\frac 12$}~\one_N + A_0
\label{covcoordsdef}\ee
which transform under the gauge group $U(N)$ as $C_\mu \mapsto
U^{-1}\, C_\mu\, U$ for $\mu=0,1,2,3$ and $U \in U(N)$. We can again
assemble them into a larger $\cN\times\cN$ matrix
\be
C = C_\mu \otimes\sigma^\mu \ . 
\label{C-collective}
\ee
Generically these are four independent fields, and we
have to somehow reduce them to two tangential fields on $S_N^2$. There
are several ways to do this. For example, one can impose the
constraints $A_0 =0$ and $C_i\, C^i = \frac{N^2-1}4~\one_\cN$ as
in~\cite{matrixsphere}, leading to a constrained hermitian multi-matrix
model describing quantum gauge theory on the fuzzy sphere which
recovers Yang-Mills theory on the classical sphere in the large $N$
limit. 

Here we will use a different approach and impose the constraints
\be
C^2 = \mbox{$\frac{N^2}4$}~\one_{\cN} \qquad \mbox{and} \qquad \Tr(C)
=N
\label{constraint}
\ee
which is equivalent to requiring that $C$ has eigenvalues $\pm\,\frac
N2$ with multiplicities $N_\pm=N\pm1$. In terms of the components of
(\ref{C-collective}), this amounts to the constraints
\be
C_i \,C^i + C_0^2=\mbox{$\frac{N^2}{4}$}~\one_{\cN} \qquad \mbox{and}
\qquad \ii\epsilon_{i}{}^{jk}\,C_j \,C_k+ \{C_0,C_i\}= 0 \ .
\label{C2}
\ee
We checked above that this is satisfied
for $A_\mu=0$, wherein $C=\Xi$. We can then consider the action of the
unitary group $U(2N)$ given by
\be
C ~\longmapsto~ U^{-1}\, C \, U
\ee
which generates a coadjoint orbit of $U(2N)$ and preserves
the constraint \eq{constraint}. The gauge fields $A_\mu$ are in this
way interpreted as fluctuations about the coordinates of the quantum
space $S_N^2$. The constraint (\ref{constraint}) ensures that the
covariant coordinates (\ref{C-collective}) describe a dynamical fuzzy
sphere. The gauge group $U(N)$ and the global isometry group $SU(2)$
of the sphere are subgroups of the larger symmetry group $U(2N)$. In
particular, the generators of the gauge group are given by elements of
the form $\phi =\phi_0 \otimes\sigma^0$.

We thus claim that a possible {configuration space of gauge
  fields} is given by the {\em single} coadjoint orbit
\be
\cO := \cO(\Xi) = \big\{ C = U^{-1}\, \Xi\, U~\big|~  U \in U(\cN\,) 
\big\}
\label{orbit-2}
\ee
where $\Xi\in \mun(2N)$ is given by (\ref{Xi-collective}). Explicitly,
dividing by the stabilizer of $\Xi$ gives a representation of the
orbit (\ref{orbit-2}) as the symmetric space $\cO\cong
U(2N)/U(N+1)\times U(N-1)$ of dimension $\dim(\cO)=2(N^2-1)$. 
Therefore the orbit $\cO$ captures the correct
number of degrees of freedom at least in the commutative limit
$N\to\infty$, where the gauge fields $A_i$ become essentially
tangent vector fields on $S_N^2$. This will be established in detail below.
A similar construction was given in~\cite{CP2paper} for the case of $\C P^2$. 

The tangent space to $\cO(\Xi)$ at a point $C$ is
isomorphic to $T_C\cO\cong\mun(\cN\,)/\mr$, where
\be
\mr=\mun(N_+)\oplus\mun(N_-)
\ee
is the stabilizer subalgebra of $\Xi$. This identification is
equivariant with respect to the natural adjoint action of the Lie
group $U(\cN\,)$. Explicitly, tangent vectors to $\cO(\Xi)$ at $C$
have the form
\be
V_\phi = \ii[C,\phi]
\label{tangentvectors}
\ee
for any hermitian element $\phi \in \mun(\cN\,)/\mr$,\footnote{With
  our conventions, the vector fields \eq{tangentvectors} are real.}
which are just the generators of the unitary group $U(\cN\,)$ acting
on $\cO(\Xi)$ by the adjoint action. These actually describe vector
fields on the entire orbit space $\cO(\Xi)$. Here and in the following
we use the symbol $C$ to denote both elements of $\cO(\Xi)$, as well
as the matrix of overcomplete coordinate functions on $\cO(\Xi)$
defined using the embeddings
$\cO(\Xi)\hookrightarrow\mun(\cN\,)\hookrightarrow\C^{\cN^2}$.

The generalization to nonabelian $U(n)$ gauge theory is very
simple. One now takes
\be
\cN = 2 n \,N
\label{cN2nN}\ee
and enlarges the matrix \eq{Xi-collective} to $\Xi\otimes\one_n$
(which we continue to denote as $\Xi$ for ease of notation). The
configuration space is given by the $U(\cN\,)$ orbit (\ref{orbit-2})
with $C^2 = \frac{N^2}4~\one_\cN$ and
\be
\Tr(C) = n\,N \ .
\label{UnTrconstr}\ee
Then $C$ has eigenvalues $\pm \,\frac N2$ of respective multiplicities
$n\,(N\pm1)$, and the configuration space
\be
\cO = U(2n\,N)/U(n\,N_+)\times U(n\,N_-)
\label{nonaborbit}
\ee
describes $\mun(n)$-valued gauge fields on $S^2_N$. Its dimension is
given by
\be
\dim(\cO) = 2 n^2\,\left(N^2-1\right) \ .
\label{dim-nonabelian}
\ee

\subsection{Yang-Mills action\label{YMAction}}

We claim that the action
\be
S=S(C) := \mbox{$\frac{N}g$}\,\Tr\big(C_0-\mbox{$\frac
  12$}~\one_{n\,N}\big)^2 
\label{YM-action}
\ee
for $C \in \cO$ reduces in the commutative limit $N \to \infty$ to the
usual Yang-Mills action on the sphere $S^2$. It can therefore be taken
as a definition of the Yang-Mills action on the fuzzy sphere
$S^2_N$. We establish this explicitly below in the abelian case
$n=1$.

Consider the three-component field strength~\cite{matrixsphere}
\bea
F_i &:=& \ii\epsilon_{i}{}^{jk}\,C_j\, C_k + C_i  \nn\\[4pt]
 &=& \ii\epsilon_{i}{}^{jk}\,[\xi_j, A_k] + 
\ii\epsilon_{i}{}^{jk}A_j\, A_k + A_i \ 
\label{fieldstrength-3}
\eea
where $C_i = \xi_i + A_i$ as in \eq{covcoordsdef}.
To understand its significance, consider the ``north pole'' of $S_N^2$
where $\xi_3 \approx \frac N2 \,x_3 = \frac N2~\one_N$ (with unit radius),
and one can replace the operators
\be
\ii\,{\rm ad}_{\xi_i} \;\;\longrightarrow \;\;
-\varepsilon_{i}{}^{j}\,\partial_j:=
-\varepsilon_{ij}\,\mbox{$\frac{\partial }{\partial x_{j}}$}
\ee
in the commutative limit for $i,j = 1,2$. Hence upon identifying the
classical gauge fields $A^{\rm cl}_i$ through
\be
A^{\rm cl}_i = -\varepsilon_{i}{}^{j}\, A_j \ ,
\ee
the ``radial'' component $F_3$ of the field strength
\eq{fieldstrength-3} reduces in the commutative limit to the standard
expression
\be
F_3 \approx \partial_1 A_{2}^{\rm cl}-\partial_2 A_{1}^{\rm cl} 
+ \ii \big[A_1^{\rm cl}\,,\,A_2^{\rm cl}\big] \ .
\ee
The constraint \eq{C2} now implies
\bea
F_i  + \big\{C_0-\mbox{$\frac 12$}~\one_N\,,\,C_i\big\} ~=~
F_i  + \big\{A_0\,,\,C_i\big\} &=& 0 \ , \nn\\[4pt]
\big\{\xi_i\,,\, A^i\big\} + A_0 + A_i\, A^i + A_0\, A_0 &=& 0 \ .
\label{constraint-F}
\eea
Since only configurations with $A_0 = O(\frac 1N)$
have finite action \eq{YM-action} and $\xi_3$ is of order $N$, this
implies that $A_3$, $F_1$ and $F_2$ are of order $\frac 1N$ at the
north pole, with  $A_{1}$ and $A_2$ finite of order $1$. In
particular, only the radial component $F_3$ survives the $N \to
\infty$ limit, with 
\be
F_3 = -\{A_0,C_3\} \approx -N\, A_0 \ .
\label{rho-F}
\ee
This analysis can be made global by considering the ``radial'' field
strength  $F_r = x^i\,  F_i$, which reduces to the usual field
strength scalar on $S^2$. The action \eq{YM-action} thus indeed 
reduces to the usual Yang-Mills action in the commutative limit
with dimensionless gauge coupling $g$, giving
\be
S \approx \frac 1{N\,g} \,\Tr (F_r)^2 \approx \frac 1{4\pi\, g}
\,\int_{S^2}\,\ddd\Omega~ (F_r)^2 \ .
\ee

\subsection{Critical surfaces}
\label{sec:crit-surfaces}

The critical surfaces of the action \eq{YM-action} are easy to find.
Since the most general variation of $C \in  \cO$ is given by
$\delta C = [C,\phi]$, the critical points satisfy
\be
0 =  \Tr\big(\delta C_0\, (C_0-\mbox{$\frac 12$})
\big) =  \Tr\big([C,\phi]\, C_0\big) = - \Tr\big(\phi\, [C, C_0]\big)
\ee
for arbitrary $\phi\in\mun(\cN\,)$. Hence they are given by solutions
of the equation
\be
[C_0,C]=0 \ .
\label{eom}
\ee
This agrees with the known saddle-points in the formulation
of~\cite{matrixsphere}. The equation $[C_0,C_i]=0$ together with
$C^2=\frac{N^2}4~\one_{\cN}$ in \eq{C2} implies that 
\bea
[C_i, C_j] &=& \ii \epsilon_{ijk} \,(2 C_0)\; C_k \ , \nn\\[4pt]
C_0^2 &=& \frac{N^2}{4} - \sum_{i=1}^3\, C_i^2 \ .
\label{critical-YM}
\eea
This means that $C_i$ generates an $SU(2)$-module $\pi_{n\,N}$ given
by a sum of irreducible representations of \eq{critical-YM}
characterized by partitions $\vec n=(n_1,\dots,n_k)$ of the integer
\be
n\,N=n_1+\dots+n_k \ ,
\ee
where $n_i\in \N$ is the dimension of the $i$-th
irreducible subrepresentation in the representation
$\pi_{n\,N}$. Therefore each critical point is labelled (up to gauge
transformations) by the set of dimensions $n_i$ of the irreducible
representations, supplemented by a ``sign'' which is defined by $s_i =
{\rm sgn}(C_0(n_i)) = \pm \,1$ (in that irreducible representation) if
$C_0(n_i) \neq 0$, and $s_i =0$ if $C_0(n_i)=0$. We denote the
collection of them by $\cC_{(n_1,s_1),\dots,(n_k,s_k)}$.

In particular, the ``classical'' saddle-points which in the 
commutative limit $N \to \infty$ go over to the saddle-points
of classical Yang-Mills theory on $S^2$ (often called instantons) 
are given by the critical surfaces $\cC_{(n_1,1),\dots,(n_n,1)}$ with 
\be
n_i = N - m_i\qquad \mbox{and} \qquad \sum_{i=1}^n\, m_i=0
\label{nidomclass}\ee
with small $m_i\in\Z$, for which
\be
C_0(n_i) = \mbox{$ \frac{N}{2(N-m_i)}~\one_{n_i} \approx \frac 12
  \,\big(1+\frac{m_i}N\big)~\one_{n_i}$} \ .
\ee
Note that then
\be
\Tr(C_0)  = \frac{n\,N}2
\label{TrC0dom}
\ee
as required. It follows that the action \eq{YM-action} evaluated on
these solutions is given by
\be
S\big((n_1,1)\,,\,\dots\,,\, (n_n,1)\big)
\approx\frac 1{4g}\, \sum_{i=1}^n\,  m_i^2 \ ,
\label{action-eval-2}
\ee
which is the usual expression~\cite{Minahan:1993tp,Gross:1994mr} for
the classical action of $U(n)$ Yang-Mills theory on the sphere $S^2$
with trivial gauge bundle evaluated on the two-dimensional instanton
on $S^2$ corresponding to a configuration of $n$ Dirac monopoles of
magnetic charges $m_i\in\Z$. Non-trivial gauge bundles over $S^2$ of
first Chern class ${\sf c}_1\in\Z$ are obtained by modifying the trace
constraint as in \cite{Steinacker:2007iq}. All other non-classical
saddle-points such as fluxons are suppressed at least by factors
$\eee^{-N\,g}$, reflecting the fact that their action becomes 
infinite in the commutative limit $N \to \infty$.

\subsection{Partition function\label{LocPrinc}}

We can now proceed to compute the partition function of
quantum Yang-Mills theory on the fuzzy sphere defined by the action
(\ref{YM-action}) on the configuration space (\ref{orbit-2}) of gauge
fields. The crucial aspect of the above formulation of $U(n)$
Yang-Mills theory on $S^2_N$ is that the space of gauge fields $\cO$
in \eq{orbit-2} or \eq{nonaborbit} is a coadjoint orbit. This implies
that it is in particular a symplectic (and even K\"ahler) space with
symplectic two-form $\omega$, which is given explicitly by the usual
Kirillov-Kostant construction
\be
\langle\omega,V_\phi\wedge V_\psi\rangle = \ii\Tr\big(C\,[\phi,\psi]
\big)
\label{symplectic-form}
\ee
where $V_\phi,V_\psi$ are tangent vectors to $\cO$ as in
\eq{tangentvectors}.

After an irrelevant shift of the covariant coordinates
\eq{covcoordsdef} which is equivalent to working with the reduced
Yang-Mills action 
\be
S' = S+\frac{n\,N^2}{4g} \ ,
\label{Sprime}
\ee
the partition function is defined by 
\bea
Z'&:=&\frac1{\vol(G)}\,\left(\frac{g}{4\pi\,N}\right)^{\dim(G)/2}\,
\int_{\cO}\, \ddd C~ \exp\Big(-\mbox{$\frac{N}{g}$}\,
\Tr\big(C_0^2\big)\Big) \nn\\[4pt] &=&
\frac1{\vol(G)}\,\left(\frac{g'}{2\pi}\right)^{\dim(G)/2}\,
\int_{\cO}\, \exp\Big(\omega -\mbox{$\frac{1}{2g'}$}\,
\Tr \big(C_0^2\big)\Big)
\label{Z-1}
\eea
where we have used the fact that the symplectic volume form
$\omega^d/d!$, with $d:=\dim_\C(\cO)$, defines the natural gauge
invariant measure on $\cO$ provided by the Cartan-Killing riemannian
volume form (up to some irrelevant normalization). This
follows from the fact that the natural invariant metric on $\cO$ is a
K\"ahler form. We have divided by the volume of the gauge group
$G=U(n\,N)$ with respect to the invariant Cartan-Killing form and by
another normalization factor for later convenience, and also
introduced the rescaled gauge coupling
\be
g' = \frac g{2N} \ .
\label{gprime}\ee
We will now describe, following~\cite{witten,szabo-paniak}, how the
technique of nonabelian localization can be applied to evaluate the
symplectic integral (\ref{Z-1}) exactly.

We begin by using a gaussian integration to rewrite \eq{Z-1} as
\be
Z' = \frac1{\vol(G)}\,\int_{\mg\times \cO}\, 
\Big[\,\frac{\ddd\phi}{2\pi}\,
\Big]~\exp\Big(\omega -\ii \Tr(C_0\, \phi) -\mbox{$
\frac{g'}{2}$}\, \Tr \big(\phi^2\big)\Big) \ ,
\label{Z-2}\ee
where the euclidean measure for integration over the gauge algebra
$\phi\in\mg=\mun(n\,N)$ is determined by the invariant Cartan-Killing
form. It is not hard to show that $H_\phi = \Tr(C_0\, \phi)$ is the
moment map for the action of the gauge group, which means that
\be
\ddd \Tr(C_0\, \phi) = -\iota_{V_\phi} \omega \ .
\label{moment-property}
\ee
Introduce the BRST operator
\be
Q= \ddd - \ii \iota_{V_\phi} \ ,
\ee
where $\ddd$ is the exterior derivative on $\Omega(\cO)$ and the
contraction $\iota_{V_\phi}$ acts trivially on $\phi$. It preserves
the gradation if one assigns charge~$+2$ to the elements $\phi$ of
$\mg$, and it satisfies
\be
Q^2 = -\ii\{\ddd,\iota_{V_\phi}\} = -\ii {\cal L}_{V_\phi}
\ee
where ${\cal L}_{V_\phi}$ is the Lie derivative along the vector field
$V_\phi$. Thus $Q^2=0$ exactly on the space
\be
\Omega_G(\cO):=\big(\C[[\mg]]\otimes\Omega(\cO)\big)^G
\ee
consisting of gauge invariant differential forms on $\cO$ which take
values in the ring of symmetric functions on the Lie algebra $\mg$.

By construction one has
\be
Q\big(\omega -\ii \Tr(C_0 \,\phi)\big) =0
\ee
using $\ddd\omega=0$ and \eq{moment-property}, and
\be
Q\Tr \big(\phi^2\big) =0 \ .
\ee
Therefore, the integrand of the partition function (\ref{Z-2}) defines
a $G$-equivariant cohomology class in $H_G(\cO)$, and the value of $Z'$
depends only on this class. The integral of any $Q$-exact equivariant
differential form in $\Omega_G(\cO)$ over $\mg\times\cO$ is
clearly~$0$, as is the integral of any $\iota_{V_\phi}$-exact form
even if its argument is not gauge invariant. Thus $Z'$ is unchanged by
adding any $Q$-exact form to the action, which will fix a gauge for
the localization. Hence we can replace it by
\be
Z' = \frac1{\vol(G)}\,\int_{\mg\times\cO}\,
\Big[\,\frac{\ddd\phi}{2\pi}\,\Big]~ 
\exp\Big(\omega -\ii \Tr(C_0\, \phi) -\mbox{$\frac{g'}{2}$}\, \Tr 
\big(\phi^2\big)+ t~ Q\alpha\Big) \ ,
\label{Z-3}
\ee
which is independent of $t\in\R$ for any $G$-invariant one-form
$\alpha$ on $\cO$, where
\be
Q\alpha = \ddd\alpha - \ii \langle\alpha,V_\phi\rangle \ .
\label{Qalpha}\ee
The independence of \eq{Z-3} on the particular representative
$\alpha\in\Omega(\cO)^G$ of its equivariant cohomology class will play
a crucial role in our evaluation of the partition function.

Expanding the integrand of (\ref{Z-3}) by writing $\exp(t~\ddd\alpha)$
as a polynomial in $t$ and using the fact that the configuration space
$\cO$ is compact, it follows 
that for $t \to\infty$ the integral localizes at the stationary points
of $\langle\alpha,V_\phi\rangle$ in $\mg\times\cO$. By writing $V_\phi
= V_a ~\phi^a$, where $\phi^a$ is an orthonormal basis of $\mg^\vee$,
we have $\langle\alpha,V_\phi\rangle = \langle\a,V_a\rangle~\phi^a$
and the critical points are thus determined by the equations
\bea
 \langle\alpha,V_a\rangle &=& 0 \ , \label{crit-1}\\[4pt]
\phi^a ~\ddd\langle\alpha,V_a\rangle &=& 0 \ . \label{crit-2}
\eea
Since \eq{crit-2} is invariant under rescaling of $\phi$ and the Lie
algebra $\mg$ is contractible, the homotopy
type of the space of solutions in $\mg\times\cO$ is unchanged by
restricting to $\phi=0$ and the saddle-points reduce to the zeroes
of $\langle\alpha,V_a\rangle$ in $\cO$.

Let us consider explicitly the invariant one-form $\alpha$ given
by~\cite{witten,szabo-paniak}
\be
\a=-\ii\Tr\big(C_0 \,[C,\ddd C]_0\big) \ .
\label{loc1form}
\ee
We claim that the vanishing locus of $\langle\alpha,V_a\rangle$ in
this case coincides with the critical surfaces of the original
Yang-Mills action (\ref{YM-action}) as found in
Section~\ref{sec:crit-surfaces}. To see this, we note that the
condition
\be
0 = \langle\alpha,V_a\rangle =\Tr\big(C_0\,[C\,,\,[C,\phi^a]\,]_0
\big)=-\Tr\big([C,C_0]\,[C,\phi^a]\big)
\ee
certainly holds whenever $[C,C_0]=0$. On the other hand, by
setting $\phi = C_0$ it implies
\be
0=\langle\alpha,V_\phi\rangle = -\Tr\big([C,C_0]^2\big)
\ee
which by nondegeneracy of the inner product defined by the trace
implies that $[C,C_0] =0$. Therefore the action in \eq{Z-3} has indeed
the same critical points as the Yang-Mills action~\eq{YM-action}.

Let us now explicitly establish, following~\cite{szabo-paniak}, the
localization of the partition function onto the classical solutions of
the gauge theory. Plugging (\ref{loc1form}) and (\ref{Qalpha}) into
\eq{Z-3} and carrying out the integration over $\phi\in\mg$ gives
\bea
Z' &=& \frac1{\vol(G)}\,\int_{\mg\times \cO} \,\Big[\,\frac{\ddd\phi}
{2\pi}\,\Big] ~\exp\Big(t~\ddd\a+\omega\Big)\nn\\ &&
\qquad\qquad \times~
\exp\Big(-\ii \Tr(C_0\, \phi) -\mbox{$\frac{g'}{2}$} \Tr
\big(\phi^2\big)- \ii t  \,\Tr\big([C\,,\,[C,C_0]\,]\,\phi\big)
\Big)\nn\\[4pt] \label{Z-4}
&=& \frac1{\vol(G)}\,\left(\frac{g'}{2\pi}\right)^{\dim(G)/2}\,\int_{\cO}\,
\exp\Big(t~\ddd\a + \omega \Big)\\ && \qquad\qquad \times~
\exp\Big( -\mbox{$\frac{1}{2g'}$}\, \Tr\big(C_0^2\big) 
+ \mbox{$\frac t{g'}$}\, \Tr\big(C_0\,[C\,,\,[C,C_0]\,]\big)
-\mbox{$\frac{t^2}{2g'}$}\, \Tr\big([C\,,\,[C,C_0]\,]\big)^2\Big)\nn
\eea
where we have used $\Tr(C\,[C,-]) =0$. The only configurations which
contribute to (\ref{Z-4}) in the large $t$ limit are therefore
solutions of the equation
\be
[C\,,\,[C,C_0]\,]=0
\ee
which implies as in~\cite{szabo-paniak} that
\be
0 = \Tr\big(C_0 \,[C\,,\,[C,C_0]\,]\big) = - \Tr\big([C,C_0]^2
\big) \ ,
\ee
giving $[C,C_0]=0$ as desired. Therefore the integral \eq{Z-4}
receives contributions only from the solutions of the Yang-Mills
equations \eq{eom}, which establishes the claimed localization.

The local geometry in $\mg\times\cO$ about each critical point
determines the partition function as a sum of local contributions
involving the values of the Yang-Mills action evaluated on the
classical solutions. This is gotten by considering an equivariant
tubular neighbourhood $\cN_{(n_1,s_1),\dots,(n_k,s_k)}$ of each
critical surface $\cC_{(n_1,s_1),\dots,(n_k,s_k)}$ in
$\mg\times\cO$. Since the partition function \eq{Z-3} is independent
of $t$, we can consider its large $t$ limit as above, and this limit
will always be implicitly assumed from now on. Let $\cW$ be a compact
subset of $\cO$ with $\cW\cap\cC=\emptyset$, where
$\cC:=\bigcup_{(n_i,s_i)}\,\cC_{(n_1,s_1),\dots,(n_k,s_k)}$. Then the
integral over $\cW$ in \eq{Z-4} has a gaussian decay in
$t\to\infty$. This means that in expanding $\exp(t~\ddd\a  + \omega)$
into a finite sum of terms of the form $\omega^p\wedge (t~\ddd \a)^{ m}$,
we can disregard all terms which contain $\omega$ since they will
be suppressed by factors of $\frac 1t$ and vanish in the large $t$
limit. The only terms which survive the $t\to\infty$ limit are those
with $p=0,m=d$, and the integral therefore vanishes unless $\omega$ is
replaced by $\ddd\alpha$, except at the saddle point where
$\ddd\alpha=0$. Then one has
\be
Z' = \frac1{\vol(G)}\,
\int_{\mg\times\cO}\, \Big[\,\frac{\ddd\phi}{2\pi}\,\Big]~
\exp\Big(t\, \big(\ddd\a - \ii \langle\a,V_\phi\rangle\big)\Big)~ 
\exp\Big(-\ii \Tr(C_0\, \phi) -\mbox{$\frac{g'}{2}$}\, \Tr \big(\phi^2
\big)\Big)
\label{Z-5}
\ee
in the vicinity of any critical point in which $\ddd\a$ is
nondegenerate.

\setcounter{equation}{0}
\section{Local geometry of the configuration space}
\label{Localgeom}

To proceed with the explicit evaluation of the contributions
from each critical surface $\cC_{(n_1,1),\dots,(n_k,1)}$ to the
partition function \eq{Z-5} for gauge theory on the fuzzy sphere,
we need to describe the local geometry of the configuration space
$\cO$ in the infinitesimal neighbourhoods
$\cN_{(n_1,s_1),\dots,(n_k,s_k)}$. This is achieved using the explicit
form of the complex structure $\cJ$ on the coadjoint orbits, together
with equivariance under global $SU(2)$ rotations.

\subsection{Complex structure} 
\label{sec:map-J}
 
Consider for fixed $ C \in\cO$ the map
\bea
\cJ\,:\, \mun(\cN\,)~ &\longrightarrow&~ \ms\mun(\cN\,) \nonumber \\
           \phi &\, \longmapsto& \,\,\mbox{$\frac1N$}\,V_\phi=
\mbox{$\frac{\ii}N$}\, \big[C\,,\,\phi\big] 
\eea
where as always the tangent space $T_C\cO$ at $C$ is viewed as a
subspace of the ambient space $\mun(\cN\,) \supset \cO$.
It is easy to check that it satisfies
\be
\cJ^3 = -\cJ \ .
\ee
The map $\cJ$ is the complex structure on $T_C \cO = {\rm im}(\cJ)$.
It provides the Cartan decomposition corresponding to the symmetric
space $ \cO$:
\be
\mun(\cN\,) ~=~ \ker(\cJ)~ \oplus~
\underbrace{\ker\big(\cJ^2+\one_\cN\big)}_{T_C \cO} \ .
\ee
Here $\mr=\ker(\cJ) =\mun(n\,N_+)\oplus\mun(n\,N_-)$ is the
stabilizer subalgebra of the coadjoint orbit $\cO$.

Now consider the map
\be
\begin{array}{cccc}
\mg \quad \longrightarrow & \cJ(\mg)        & \longrightarrow    &
\cJ^2(\mg) \\ & \mbox{\small (pure gauge)} &  & \mbox{\small
  (physical)}
\end{array}
\ee
which defines subspaces $\cJ(\mg), \cJ^2(\mg)$ of $T_C \cO$. We denote
with $\ms \subset \mg$ the stabilizer of the gauge transformations,
with $\cJ(\ms)=0$. One can show~\cite{Steinacker:2007iq} that
for the vacuum solution $ C =\Xi=\frac 12~\one_\cN + \xi_i \otimes
\one_n\otimes\sigma^i$, there is a splitting
\be
T_C \cO  = \cJ(\mg) \oplus \cJ^2(\mg)  
\label{tangentsplit}\ee
while the structure of the generic critical surfaces is somewhat more
complicated: 
\be
\cJ(\mg\ominus\mh) \oplus \cJ^2(\mg\ominus\mh)
\oplus E_0 \oplus E_1 ~=~ T_C\cO   
\label{tangent-decomp}
\ee
where the subalgebra $\mh$ is defined via
\be
 E_0 =\cJ(\mg) \cap \cJ^2(\mg) =  \cJ(\mh) =\cJ^2(\mh) 
\ee
and $E_1$ is an extra vector space. To determine the vector spaces
$E_0,E_1$ explicitly, we need to describe the decomposition under the
global $SU(2)$ symmetry. We will only sketch the resulting structure
and refer the interested reader to~\cite{Steinacker:2007iq} for
details.

\subsection{$SU(2)$-equivariant decomposition at critical surfaces}

The critical surface $\cC_{(n_1,1),\dots, (n_n,1)}$ defines
$ SU(2)$ generators
\be
J_i = \frac{C_i}{2 C_0} + \frac 12\, \sigma_i \qquad \mbox{with}
\quad [J_i, C] =0  \ .
\ee
We decompose everything under this action of $ SU(2)$. For the
$n\,N$-dimensional representation $\pi_{n\,N}:\mathfrak{su}(2)\to{\rm
  End}(V)$ one has
\be
V \otimes \C^2 = \Big(\,\bigoplus_{i=1}^n\, (n_i+1)\Big)~ 
 \oplus~ \Big(\,\bigoplus_{i=1}^n\,(n_i-1)\Big) 
\ee
so that
\be
C = \frac N2\,
\left(\begin{array}{cc}\mbox{$\bigoplus\limits_{i=1}^n$}\, 
\one_{(n_i+1)} & 0
      \\ 0 & -\,\mbox{$\bigoplus\limits_{i=1}^n$}\,
      \one_{(n_i-1)} \end{array}\right) ~ \subset~ \mun(\cN\,) 
\ee
and
\be
 T_C\cO \cong \Big\{\left(\begin{array}{cc}0 & X \\
                           X^\dagger & 0\end{array}\right)~\Big|~
X\in{\rm Mat}_{n\,N}\Big\} ~\subset~ \mun(\cN\,) \ .
\ee
It follows that
\be
T_C\cO\cong\bigoplus_{i,j=1}^n\, (n_i+1)\otimes (n_j-1)
\ee
and
\be
\mg\cong\bigoplus_{i,j=1}^n\, (n_i) \otimes (n_j) \ .
\ee
All this allows for the explicit computation of $ \cJ, E_0, E_1 $ for
the various critical surfaces $\cC_{(n_1,1),\dots, (n_n,1)}$ as follows:

\begin{enumerate}
\item[1)] {\sl Vacuum surface $\cC_{(N,1),\dots, (N,1)}$}

The vacuum surface is the orbit through
$C =\Xi=\frac12~\one_\cN+\xi_i \otimes \one_n \otimes \sigma^i$. The 
stabilizer is given by $\ms = \mun(n)\subset \mg$, and
\be
\mg \cong \big((1) ~\oplus~ (N+1) \otimes (N-1)\big)\otimes \mun(n) 
\ee
gives
\be T_C\cO = \cJ(\mg) \oplus \cJ^2(\mg)
\label{tangentsplit2}\ee
as in (\ref{tangentsplit}), which can be seen even by just counting
dimensions.

\item[2)] {\sl Maximally non-degenerate critical surface
    $\cC_{(n_1,1),\dots, (n_n,1)}$}

For a generic critical surface $\cC_{(n_1,1),\dots, (n_n,1)}$ with
$n_1 > n_2 > \dots > n_n$ one finds
\be
\mg \cong \bigoplus_{i,j=1}^n\, (n_i) \otimes (n_j) 
= \bigoplus_{i,j=1}^n\,  \big( (|n_i-n_j|+1) 
\oplus \cdots\oplus(n_i+n_j-1)\big) \ .
\ee
This gives
\be
T_C\cO = \cJ(\mg\ominus\mh) \oplus \cJ^2(\mg\ominus\mh)
\oplus E_0 \oplus E_1 
\label{tangent-decomp2}\ee
as in (\ref{tangent-decomp}) with 
\be
E_1 = 
\bigoplus_{i>j}\,\big(|n_i-n_j|-1\big)
\qquad \mbox{and} \qquad E_0 = \bigoplus_{i>j}\,\big(|n_i-n_j|+1\big)
\ .
\ee
\end{enumerate}

\setcounter{equation}{0}
\section{Nonabelian localization at the vacuum surface}

We will first consider the localization of the partition
function \eq{Z-5} at the vacuum orbit
\be
\cO_0 := \cC_{(N,1),\dots,(N,1)} 
= \big\{g\, \Xi \,g^{-1}~\big|~ g \in G\big\}
\cong G/U(n) ~\subset~ \cO 
\ee
with gauge group $G =U(n\,N)$ and stabilizer $ \ms = \mun(n)$.

\subsection{Statement of result}

\begin{theorem}\label{vacpartthm}
The contribution to the quantum partition function for $U(n)$
Yang-Mills theory on $S_N^2$ from the vacuum moduli space $\cO_0$ is
given by
\be
Z_0 = \frac{1}{n!}\,\frac 1{(2\pi)^{n^2+n}}~\,
\int_{\R^n}\, [\ddd s]~\Delta(s)^2~
\e^{ -\frac{g}{4}\, \sum_i\, s_i^2} \ .
\label{Z0final}
\ee
\end{theorem}

Here
\be
\Delta(s)=\prod_{i<j}\,(s_i-s_j)=\det_{1\leq i,j\leq n}\,
\big(s_i^{j-1}\big)
\ee
is the Vandermonde determinant, and we substituted back the original
Yang-Mills action $S$ using the shift \eq{Sprime}. The quantum
fluctuation integral (\ref{Z0final}) is the standard
expression~\cite{Minahan:1993tp} for the contribution from the global
minimum of the Yang-Mills action on $S^2$ to the $U(n)$ sphere
partition function. It arises from the trivial instanton configuration
with vanishing monopole charges $m_i=0$ in \eq{nidomclass}.

\subsection{Proof of Theorem~\ref{vacpartthm}}

Localization implies that we can restrict ourselves to a
$G$-equivariant tubular neighbourhood $\cN_0=\cN_{(N,1),\dots,(N,1)}$
of the critical surface, under the action of the gauge group $G =
U(n\,N)$. The neighbourhood $\cN_0$ has an equivariant
retraction by a local equivariant
symplectomorphism onto the {\it local symplectic model} ${\cal F}_0$.
 This means that the tangent space to
${\cal F}_0$ at the vacuum critical point $C$  is
given by $T_C \cO_0 \oplus\cJ^2(\mg\ominus\ms)
\cong\cJ(\mg\ominus\ms)\oplus\cJ^2(\mg\ominus\ms) = T_C \cO$, and the
symplectic two-form on ${\cal F}_0$ is simply $\omega$. 
In physical terms, the gauge fields are decomposed along the
vacuum moduli space $\cO_0$ plus infinitesimal non-gauge variations in
the subspace $\cJ^2(\mg\ominus\ms)$.

We need to introduce an explicit basis of $T_C\cO$
and of the dual space of one-forms $\Omega^1(\cO)$.  
According to (\ref{tangentsplit2}), we take the basis of vector fields
to be
\be
J_i = \cJ(g'_i\,) \ , \quad  \tilde J_j = \cJ^2(g'_j) ~\in~ T_C(\cO)
\label{JJtilde-def}
\ee
where $g'_i $ is an orthonormal basis of $\mg\ominus\ms$.
The dual basis of one-forms 
\be
\lambda^i \ , \quad  \tilde \lambda^j ~\in~ \Omega^1(\cO)
\ee
satisfy
\be
\big\langle\lambda^i\,,\, J_j\big\rangle = \delta^i{}_j \ ,
\quad \big\langle\,\tilde\lambda^i\,,\, \tilde J_j\big\rangle =
\delta^i{}_j \qquad \mbox{and} \qquad \big\langle\lambda^i\,,\,\tilde
J_j\big\rangle =\big\langle\,\tilde\lambda^i\,,\, J_j\big\rangle = 0 \ .
\ee
Now introduce functions $f_i = \langle\a,J_i\rangle$.
One can show that $\langle\a, \cJ^2(\mg)\rangle = 0$, which
implies that the localization one-form can be expanded as
\be
\a =  f_i~ \lambda^i
\ee
with
\be
\ddd\a = \ddd f_i\wedge\lambda^i + f_i ~\ddd\lambda^i \ .
\ee
In particular, one has
\be
\frac{(\ddd\a)^d}{d!} =
\bigwedge_{i=1}^d\, \left(\ddd f_i \wedge \lambda^i\right)
\ee
up to forms which  vanish on-shell, and hence are
killed by localization in the large $t$ limit. Here
$d=\dim_\C(\cO)=n^2\,(N^2-1)$ is the (real) dimension of the
vacuum orbit $\cO_0$. 

We can now proceed with the evaluation of 
the local contribution to the partition function
(\ref{Z-5}) for $t\to\infty$:
\bea
Z'_0 &=&\frac1{\vol(G)}\,
\int_{\mg\times {\cal F}_0}\,\Big[\,\frac{\ddd\phi}{2\pi}\,\Big]~
\frac{t^d}{d!}\,(\ddd\a)^{d}~
\e^{-\ii t\,\langle\a,V_\phi\rangle-\ii \Tr(C_0\, \phi) -\frac{g'}{2}\,
\Tr (\phi^2)}\nn\\[4pt]
&=& \frac1{\vol(G)}\,
\int_{\mg\times {\cal F}_0}\,\Big[\,\frac{\ddd\phi}{2\pi}\,\Big]~
t^d~\bigwedge_{i=1}^d\, \left(\ddd f_i \wedge \lambda^i\right)~
\e^{-\ii N\,t\, f_i \,\phi^i-\ii \Tr(C_0\, \phi) -\frac{g'}{2} \,
\Tr (\phi^2)} \nn\\[4pt]
&=& \frac1{\vol(G)}\,
\int_{\ms}\,\Big[\,\frac{\ddd\phi}{2\pi}\,\Big]~
\e^{-\ii \Tr(C_0\, \phi) -\frac{g'}{2}\,\Tr (\phi^2)}~
\frac 1{N^{d}}\,\int_{\cO_0}\, \bigwedge_{i=1}^{d}\,\lambda^i \ .
\label{Z0-1}\eea
Here the $f_i$ integrals over the fibre $\cJ^2(\mg\ominus\ms)$
have produced delta-functions setting $\phi^i=0$ in
$\mg\ominus\ms$. We can carry out the integral over the moduli space
$\cO_0$ in \eq{Z0-1} by observing that
\be
\frac 1{N^{d}}\,\int_{\cO_0}\,\bigwedge_{i=1}^{d}\,\lambda^i=\int_{G/S}\,
\bigwedge_{i=1}^{d}\,\eta^i= \frac{{\rm vol}(G)}{{\rm vol}(S)} \ ,
\label{intcO0}
\ee
where the pullbacks $\cJ^*(\lambda^i) = \eta^i$ define left-invariant
one-forms on the gauge group $G$.

To evaluate the remaining integral over the gauge stabilizer algebra
$\ms\cong\mun(n)$ in \eq{Z0-1}, we note that
 the integrand defines a gauge
invariant function $f:\mun(n)\to\R$. It can therefore be written using
the Weyl integral formula as 
\be
\int_{\mun(n)}\,[\ddd\phi]~f(\phi)=\frac{{\rm vol}\big(U(n)\big)}
{n!\,(2\pi)^n}\,\int_{\R^n}\,[\ddd s]~\Delta(s)^2~f(s) \ ,
\label{Weylint}
\ee
where the Vandermonde determinant is the Weyl determinant for
$U(n)$ arising as the jacobian for the diagonalization of hermitian
matrices on the left-hand side of (\ref{Weylint}). 
{}From \eq{Z0-1}--\eq{Weylint} we obtain 
\bea
Z'_0 &=& \frac1{\vol(S)}\, \int_{\ms}\,
\Big[\,\frac{\ddd\phi}{2\pi}\,\Big]~
\e^{-\ii \Tr(C_0\, \phi) -\frac{g'}{2}\,\Tr (\phi^2)} \nn\\[4pt]
&=&\frac{1}{n!}\, \frac 1{(2\pi)^{n^2}}\,
\int_{\R^n}\, \Big[\,\frac{\ddd s}{2\pi}\,\Big]~
\Delta(s)^2~\e^{-\ii \frac N2\, \sum_i\, s_i  -\frac{g}{4}\,
\sum_i\, s_i^2}
\label{Z0-2}
\eea
where we used $\vol(S) =
{N}^{N^2/2}~\vol(U(n))$ with respect to the Cartan-Killing metric on
$\ms$, since $S = U(n) \otimes \one_N$. Applying the integral identity
\bea
&& \int_{\R^n}\, [\ddd s]~\Delta(s)^2~
\e^{-\ii \frac N2\, \sum_i\, s_i + \frac\ii4\, \sum_i\, m_i\, s_i
 -\frac{g}{4}\, \sum_i\, s_i^2} \nn\\
&& \qquad\qquad\qquad~=~ \e^{-\frac{n\,N^2-m\,N}{4g}}\,
\int_{\R^n}\,[ \ddd s]~\Delta(s)^2~
\e^{\frac\ii4\, \sum_i\, m_i\,s_i -\frac{g}{4}\, \sum_i\, s_i^2} 
\label{Lemma}\eea
where  $m = \sum_i\, m_i$ allows us to finally write the partition
function as in (\ref{Z0final}).

\setcounter{equation}{0}
\section{Nonabelian localization at maximally irreducible saddle
  points\label{LocMaxNon-Deg}}

We now turn to the opposite extreme and look at the local contribution
to the partition function \eq{Z-5} from a generic maximally
non-degenerate critical surface. We denote this gauge orbit by
\be
\cO_{\rm max}(\vec n):= \cC_{(n_1,1),\dots,(n_n,1)} 
= \big\{g\, C \,g^{-1}~\big|~ g \in U(n\,N-{\sf c}_1)\big\}\cong
U(n\,N-{\sf c}_1)/U(1)^n
\ee
and assume that the integers $n_1 > n_2 > \cdots > n_n$ are explicitly
specified. Here we allow also ${\sf c}_1 \neq 0$ which describes sectors 
with non-vanishing $U(1)$ monopole number.

\subsection{Statement of result}

\begin{theorem}\label{maxdegthm}
The contribution to the quantum partition function for $U(n)$
Yang-Mills theory on $S_N^2$ from a maximally non-degenerate moduli
space $\cO_{\rm max}(\vec n)$ is given by
\be
Z_{\rm max}=\frac{(-1)^{n\,(n-1)/2}~\e^{n\,N^2/4g}}
{(2\pi)^{n^2+n}}\,
\int_{\R^n}\, [\ddd s]~ \Delta(s)^{2}~
\e^{-\frac\ii2\,  N\,\sum_i\, s_i  -\frac{g}{4}\, \sum_i
  \frac{n_i}{N}\, s_i^2 } \ .
\label{Zmax-s2}\ee
\end{theorem}

Setting $\tilde s_i := \sqrt{n_i/N} \,s_i$ in (\ref{Zmax-s2}), we get
\be
Z_{\rm max}=\frac{(-1)^{n\,(n-1)/2}}{(2\pi)^{n^2+n}}\,
\frac{N^{n/2}~\e^{n\,N^2/4g}}
{\prod\limits_{k=1}^n\,\sqrt{n_k}}\,
\int_{\R^n}\, [\ddd\tilde s\,]~ \prod_{k > l} \,
\Big(\,\sqrt{\mbox{$\frac{N}{n_k}$}}\, \tilde s_k-\sqrt{
\mbox{$\frac{N}{n_l}$}}\,\tilde s_l\Big)^{2}~
\e^{-\frac\ii2\, \sum_i\, \sqrt{\frac{N^3}{n_i}}\,\tilde s_i 
-\frac{g}{4}\, \sum_i\, \tilde s_i^{\,2} } \ .
\label{Zmax-s2a}\ee
Completing the square of the gaussian function of $\tilde s_i$ in
\eq{Zmax-s2a} identifies the Boltzmann weight of the action \eq{Sprime}
on the non-degenerate solution space $\cC_{(n_1,1),\dots,(n_n,1)}$. In
the large $N$ limit, we substitute \eq{nidomclass} with $\tilde s_i
\approx\big(1+\frac{m_i}{2N}\big)\, s_i$. Neglecting terms of order
$\frac 1N$ then reduces \eq{Zmax-s2a} to
\be
Z_{\rm max}\approx \pm\, \frac{\e^{n\,N^2/4g}}{(2\pi)^{n^2+n}}\, 
\int_{\R^n}\, [\ddd s]~ \Delta(s)^{2}~
\e^{- \frac\ii2\,N \,\sum_i\, s_i}~
\e^{\frac\ii4\, \sum_i\, m_i\, s_i  -\frac{g}{4}\, \sum_i\, s_i^2 } \
,
\label{Zmax-s3}\ee
and an application of the integral identity (\ref{Lemma}) leads to the
result
\be
Z_{\rm max}\approx \pm\, \frac 1{(2\pi)^{n^2+n}}\, 
\int_{\R^n}\, [\ddd s]~ \Delta(s)^{2}~
\e^{\frac\ii4\, \sum_i\, m_i\, s_i  -\frac{g}{4}\, \sum_i\, s_i^2 } \
.
\label{Zmaxfinal}\ee
This can easily be generalized~\cite{Steinacker:2007iq} to non-trivial
$U(1)$ monopole number, or Chern class ${\sf c}_1=m=\sum_i\,m_i$. 
The form \eq{Zmaxfinal} coincides with the classical
result~\cite{Minahan:1993tp} for the contribution to the $U(n)$ sphere
partition function from the Yang-Mills instanton on $S^2$ specified by
the configuration of magnetic monopole charges
$m_1,\dots,m_n\in\Z$. In particular, using the standard manipulation
of~\cite{Minahan:1993tp} one can change integration variables in
\eq{Zmaxfinal} to identify the anticipated Boltzmann weight of the
action \eq{action-eval-2}.

\subsection{Proof of Theorem~\ref{maxdegthm}}

We want to compute the integral $Z'_{\rm max}$ in \eq{Z-5}
over a local neighbourhood $\cN_{\rm max}$ of $\cO_{\rm max}(\vec n)$,
which is independent of $t$ in the large $t$ limit. This is similar in
spirit but technically more involved than for the vacuum surface.
We first need to find a suitable basis for the tangent space $ T_{C}
\cO$ at the irreducible critical point $C$, using the splitting
(\ref{tangent-decomp2}). The definition of the basis $J_i,\tilde J_i$
introduced in \eq{JJtilde-def} naturally extends to include the
non-trivial subspaces $E_0, E_1$ in this case with
\be
J_i = \cJ(g'_i\,) \ , \quad \tilde J_j = \cJ^2(g'_j) \ , \quad
H_i = \cJ(h'_i\,) ~\in~ \cJ(\mh) = E_0 \qquad \mbox{and} \qquad K_i \in
E_1 \ ,
\ee
for $g'_i$ and $h'_i$ an orthonormal basis of
$\mg\ominus\mh\ominus\ms$ and of $\mh \ominus \ms$, respectively.
We define again $\langle\a, \cJ(g'_i\,)\rangle =f_i$.
The elements $K_i$ are assumed to form an orthonormal basis of $E_1$,
orthogonal to $\cJ(\mg) \oplus \cJ^2(\mg)$.

$E_0$ and $E_1$ are naturally complex vector spaces, whose generators
are embedded into the tangent space decomposition \eq{tangent-decomp2}
as
\be
K_{i} = \left(\begin{array}{ccccc}0&0 &\vline & 0 & 0 \\
                                      0&0 &\vline& X_{i} & 0 \\
\hline
                           0 & X_{i}^\dagger&\vline & 0&0  \\
                           0 & 0&\vline & 0&0 
\end{array}\right)
\label{E_1-explicit-2}
\ee
and similarly for $H_i$. The complex structure is given by the map
$\cJ$,  which amounts to multiplying $X_i$ by $\ii$. We accordingly
take the real basis $K_i$ to be ordered as $\{K_i\} = \{(\tilde K_i,
\cJ(\tilde K_i))\}$, and similarly for $H_i$. As matrices, all of the
generators $H_i, K_j$ are hermitian. The corresponding dual one-forms
$\beta^i,\gamma^i$ are defined as usual by
\be
\big\langle \b^i\,,\, H_j\big\rangle = \delta^i{}_j \qquad \mbox{and}
\qquad \big\langle \gamma^i\,,\, K_j\big\rangle = \delta^i{}_j
\ee
with all other pairings equal to~$0$.
One can show \cite{Steinacker:2007iq} that
\be
\ddd\a = \ddd f_i\wedge \lambda^i 
+ \mbox{$\frac 12$} \, A_{ij}~ \gamma^i \wedge \gamma^j + O_f
\ee
where $O_f$ denotes contributions which vanish on-shell, and
\be
A_{ij} =  2\ii \Tr\big(K_i~{\rm ad}_{C_0}(K_j)\big)
\label{Aijdef}\ee
is an antisymmetric matrix. One then has
\be
\frac{(\ddd\a)^{d-d_0}}{(d-d_0)!} = 
\pfaff( A)~\Big(\,\bigwedge_{i=1}^{2d_1}\,\gamma^i\Big)~ \wedge~
\Big(\,\bigwedge_{j=1}^{d-d_0-d_1}\,\ddd f_j \wedge \lambda^j\Big)
+ O_f
\ee
where $d_0$ (resp. $d_1$) is the complex dimension of the vector space
$E_0$ (resp. $E_1$), and
\be
\pfaff(A)= \epsilon^{i_1\cdots i_{2d_1}}\, 
A_{i_1 i_2}\cdots A_{i_{2d_1-1} i_{2d_1}}
\ee
is the pfaffian of the antisymmetric matrix $A=(A_{ij})$.

To proceed with the localization, we need to find the local geometry
and define its symplectic model. The $G$-equivariant tubular
neighbourhood $\cN_{\rm max}$ of $\cO_{\rm max}(\vec n)$ has an
equivariant retraction (by a local equivariant symplectomorphism) onto
the local symplectic model ${\cal F}_{\rm max}$, defined to be an
equivariant symplectic vector bundle over $\cO_{\rm max}(\vec n)$ with
fibre $\cJ^2(\mg\ominus\mh\ominus\ms)\oplus E_1$ which is a sub-bundle
of the tangent bundle $T \cO$ restricted to $\cO_{\rm max}(\vec
n)$. Thus the tangent space to ${\cal F}_{\rm max}$ is given by
\be
T_C \cO_{\rm max}(\vec n)
\oplus \cJ^2(\mg\ominus\mh\ominus\ms)\oplus E_1 ~\cong~ E_0 \oplus
\cJ(\mg\ominus\mh\ominus\ms) \oplus\cJ^2(\mg\ominus\mh\ominus\ms)\oplus
E_1~ = ~T_C \cO \ ,
\label{TF-max}
\ee
and the symplectic two-form on ${\cal F}_{\rm max}$ is simply
$\omega$. In physical terms, the gauge fields are split along
the moduli space $\cO_{\rm max}(\vec n)$, plus infinitesimal non-gauge
variations belonging to $\cJ^2(\mg\ominus\mh\ominus\ms)$ along with
unstable modes in the subspace $E_1$. Due to the presence of the
localization form $\a$ in the action, we can restrict ourselves to
this model ${\cal F}_{\rm max}$ replacing $\cN_{\rm max}$. Identically
to the case of the vacuum surface in the previous section, the
canonical symplectic integral over $\mg\times\cN_{\rm max}$ will in
this way reduce to an integral over $\ms\times\cO_{\rm max}(\vec n)$.

We may now proceed to calculate
\bea
Z'_{\rm max} &=& \frac1{\vol(G)}\,\int_{\mg\times \cN_{\rm max}}\,
\Big[\,\frac{\ddd\phi}{2\pi}\,\Big]~
\exp\Big(\omega + t \,\big(\ddd\a-\ii\langle\a,V_\phi\rangle
\big)-\ii \Tr(C_0\, \phi) -\mbox{$\frac{g'}{2}$}\, \Tr \big(\phi^2
\big)\Big) \nn\\[4pt]
&=&\frac1{\vol(G)}\,
\int_{\mg\times \cO_{\rm max}(\vec n) \times \cJ^2(\mg\ominus\mh\ominus\ms)
\times E_1} \,
\Big[\,\frac{\ddd\phi}{2\pi}\,\Big]~ \frac{(t~\ddd\a)^{d-d_0}}
{(d-d_0)!}\wedge\frac{\omega^{d_0}}{d_0!} \nn\\ && \qquad\qquad
\times~\e^{-\ii t\,\langle\a,V_\phi\rangle-\ii \Tr(C_0 \,\phi) -
\mbox{$\frac{g'}{2}$}\, \Tr(\phi^2)} \nn\\[4pt]
 &=& \frac1{\vol(G)}\,
\int_{(\mg\ominus\mh\ominus\ms) \oplus \mh \oplus \ms}\,
\Big[\,\frac{\ddd\phi}{2\pi}\,\Big]~\pfaff (A)~
\nn\\ && \qquad\qquad
\times~\int_{\cO_{\rm max}(\vec n) \times \cJ^2(\mg\ominus\mh\ominus\ms)
\times E_1}\,t^{d-d_0}~\Big(\,\bigwedge_{i=1}^{2d_1}\,\gamma^i
\Big)~ \wedge~\Big(\,\bigwedge_{j=1}^{d-d_0-d_1}\,\ddd f_j
\wedge \lambda^j\Big)~\wedge~\frac{\omega^{d_0}}{d_0!}
\nn\\ && \qquad\qquad
\times~\e^{-\ii t\, (N \,f_i\,\phi^i+ \langle \a,V_{\phi'}\rangle)
-\ii \Tr(C_0\, \phi) -\frac{g'}{2}\, \Tr (\phi^2)}
\label{Zmaxproceed}\eea
with $\phi'\in\mh\oplus\ms$. In the second line we have used the fact
that $\ddd\a$ vanishes when evaluated on the subspace $E_0$, and
therefore we need $d_0$ powers of $\omega$ to yield a non-trivial
volume form. Then $(t~\ddd\a)^{d-d_0}\wedge\omega^{d_0}$ is the only
term which survives in the large $t$ limit. We will modify this below
by adding a second localization form $\a'$ in order to write the
localization integral in the generic form (\ref{Z-5}) without the
symplectic two-form $\omega$.

We can now evaluate the integrals in \eq{Zmaxproceed} over $f_i$ in
the fibre $\cJ^2(\mg\ominus\mh\ominus\ms)$ and
$\phi^i\in\mg\ominus\mh\ominus\ms$ as in the previous section, 
which localizes for $t\to\infty$ to an integral over the
subspace $E_1$ and the gauge orbit $\cO_{\rm max}(\vec n)$ given by
\bea
Z'_{\rm max} &=&\frac1{\vol(G)}\,\int_{\mh \oplus \ms}\,
\Big[\,\frac{\ddd\phi}{2\pi}\,\Big]~
\frac{\pfaff (A)}{N^{d-d_0-d_1}}~\int_{\cO_{\rm max}(\vec n) \times E_1}\,
t^{d_1}~\Big(\,\bigwedge_{i=1}^{2d_1}\,\gamma^i\Big)~ \wedge~
\Big(\,\bigwedge_{j=1}^{d-d_0-d_1}\,\lambda^j\Big)~\wedge~
\frac{\omega^{d_0}}{d_0!}\nn\\ && 
\qquad\qquad\qquad\qquad\qquad\qquad\qquad
\times~\e^{-\ii t\,\langle \a,V_{\phi}\rangle
-\ii \Tr(C_0\, \phi) -\frac{g'}{2}\, \Tr (\phi^2)} \ .
\label{Zdeg-4}
\eea
The gauge invariant volume form for the integration domain whose
tangent space is $E_0$ is given by the symplectic volume form
$\omega^{d_0}/d_0!$, since $\ddd\a$ vanishes on $E_0$, but this will
be modified below. It remains to compute the integral over $E_1$. Upon
evaluating $\langle \a,V_\phi\rangle$ at second order on $E_1$,
i.e. away from the critical surface, we will find below that this
pairing becomes a quadratic form which leads to a localization through
a gaussian integral. However, to evaluate it explicitly it is easier
to first localize the integral over $E_0$, which presently is a
complicated non-gaussian integral which does not admit a gaussian
approximation at $t\to\infty$ and is difficult to evaluate in a closed
analytic form. But this can be done by adapting a trick  taken
from~\cite{beasley-witten}, which amounts to adding a further suitable
localization one-form $\a'$, or equivalently a cohomologically trivial
form $Q\a'$, to the action in \eq{Z-5}. Indeed, we may compute $Z'_{\rm
  max}$ using any other invariant form $\alpha'$ which is homotopic to
$\alpha$ on the open neighbourhood $\cN_{\rm max}$. The one-form
$\alpha'$ need only be non-vanishing on $E_0\subset\cN_{\rm max}$, as the
other integrals can be directly carried out.

In order to evaluate the integrals over $E_0$ and $\mh$,
following~\cite{beasley-witten} we introduce an additional
localization term $\exp(t~Q\a'\,)$ in the partition function with
\be
\a' := -\mbox{$\frac{2}N$}\,
\cJ~\ddd\Tr(C\, \phi)\Big|_{E_0} \ .
\label{alphaprime}
\ee
The projection onto $E_0$ is equivalent to projecting $\phi\in\mg$
onto $\mh$. This one-form is equivariant on-shell, and it can be
extended to the $G$-equivariant tubular neighbourhood $\cN_{\rm max}$ of
the critical surface $\cO_{\rm max}(\vec n)$ as follows.
On the tangent space $\cJ(\mg\ominus\mh\ominus\ms) \oplus E_0$ of
$T\cO_{\rm max}(\vec n)$ in \eq{TF-max} there is an equivariant
projection onto the subspace $E_0$. In this way $\a'$ is properly
defined on the local model, and can hence be extended to $\cN_{\rm
  max}$. One could also define $\a'=\frac{2\ii}N\,\chi\,
\cJ~\ddd\Tr(C\,\phi)\big|_{E_0}$ using a smooth
$G$-invariant cutoff function $\chi$ with support near the given
saddle-point and $\chi=1$ in the tubular neighbourhood, which is
globally well-defined over $\cN_{\rm max}$ as an equivariant
differential form. Note that $t_1\,\a+t_2\,\a'$ vanishes only on the
original critical points for any $t_1,t_2\in\R$ with $t_1\neq0$, and
no new ones are introduced. Then our previous computation \eq{Z-4}
would essentially go through, since $\a'$ vanishes on
$\cJ(\mg\ominus\mh\ominus\ms)$ and there are no critical points where
$\ddd\chi\neq0$. It is therefore just as good a localization form to
use as $\a$ is. It follows that the modification of the canonical
symplectic integral over $\cN_{\rm max}$ given by
\be
Z'_{\rm max} = \frac1{\vol(G)}\,\int_{\mg\times \cN_{\rm max}}\,\Big[\,
\frac{\ddd\phi}{2\pi}\Big]~\exp\Big(\omega + t_1~Q\a + t_2~Q\a'-
\ii \Tr(C_0\, \phi) -\mbox{$\frac{g'}{2}$}\,\Tr\big(\phi^2\big)\Big)
\label{Zmaxt1t2}\ee
is independent of both $t_1,t_2\in\R$. 
Then $\a'$ will localize the integral over
$\mh\subset\mg$ as well as the integral over the unstable modes in
$E_1$, without the need to expand $\langle\a,V_\phi\rangle$ to higher
order.

Let us first integrate over $\mh$. One can
show~\cite{Steinacker:2007iq} that the new localization form $\a'$
satisfies
\be
\langle \a',V_{h_i}\rangle = 2\Tr\big(H_i\,\cJ(\phi)\big) \ ,
\ee
with $h_i$ a basis of $\mh$. This produces a gaussian integral
localizing $\mh$ to the gauge stabilizer algebra
$\ms\cong\mun(1)^n$. Then
\be
\ddd\a' =\mbox{$\frac{2\ii}{N^2}$}\,\tilde A_{ij}~\b^i\wedge \b^j 
\qquad \mbox{and} \qquad \frac{(\ddd\a'\,)^{d_0}}{d_0!} 
= \left(\mbox{$\frac{4\ii}{N^2}$}\right)^{d_0}~\pfaff\big(\tilde A\,\big)~
\bigwedge_{i=1}^{2d_0}\, \beta^i \ ,
\label{dalphaprime}
\ee
where $\tilde A_{ij}:=\Tr(H_i\,[s,H_j])$ is an antisymmetric matrix
and we have restricted to $\phi=s \in \ms$ using the localization (see
\eq{locmhms} below).  Using the explicit description of the local
geometry given in Section~\ref{Localgeom}, one finds
\be
\pfaff\big(\tilde A\,\big) = 
(-\ii)^{d_0}\sqrt{\det(M)}\; \prod_{k > l} \,(s_k-s_l)^{|n_k-n_l|+1}
\label{PfaffA-eval}
\ee
where $M_{ij} := 2\Tr(H_i\, H_j)$ is a symmetric matrix. We can now
evaluate the localization integral
\be
\int_{\mh}\,\Big[\,\frac{\ddd\phi}{2\pi}\,\Big]~t_2^{d_0}\,
\frac{(\ddd \a'\,)^{d_0}}{d_0!}~\e^{-\ii t_2 \,\langle \a',V_\phi\rangle } 
=\left(\mbox{$\frac{4\ii}{N^2}$}\right)^{d_0}\,
\int_{\mh}\,\Big[\,\frac{\ddd\phi}{2\pi}\,\Big]~t_2^{d_0}~
\pfaff\big(\tilde A\,
\big)~\e^{-2\ii t_2\, \phi^{i}\, M_{ij}\, \phi^{j}} ~
\bigwedge_{i=1}^{2d_0}\, \beta^i
\ee
where $\phi = \phi^{i}\, h_{i}$. The oscillatory gaussian integral is
defined by analytic continuation $t_2\to t_2-\ii \varepsilon$ for a
small positive parameter $\varepsilon$, which we are free to do as the
partition function is formally independent of $t_2$. With this
continuation understood and a suitable orientation of the vector space
$\mh$, we readily compute
\bea
\int_{\mh}\,\Big[\,\frac{\ddd\phi}{2\pi}\,\Big]~t_2^{d_0}\,
\frac{(\ddd \a'\,)^{d_0}}{d_0!}~\e^{-\ii t_2 \,\langle \a',V_\phi\rangle } 
&=& \left(\mbox{$\frac{4\ii}{N^2}$}\right)^{d_0}\,
\left(\mbox{$\frac 1{2\pi}$}\right)^{2d_0}\,
\left(-\mbox{$\frac{\pi}{2\ii}$}\right)^{d_0}~
 \frac{\pfaff\big(\tilde A\,\big)}{\sqrt{\det(M)}}~
\bigwedge_{i=1}^{2d_0}\, \beta^i \nn\\[4pt]
&=& \frac{\ii^{d_0}}{( 2\pi\, N^2)^{d_0}}~
\prod_{k > l}\, (s_k-s_l)^{|n_k-n_l|+1}~
\bigwedge_{i=1}^{2d_0}\, \beta^i \ .
\label{locmhms}
\eea
This integral thus produces a measure on $\ms$ which we will use below
to perform the remaining integral over the stabilizer.

Now that the $\phi$-integration in \eq{Zdeg-4} is localized onto
$\ms$, we can proceed to evaluate the integral over $E_1$. This space
has a basis $K_i$ as introduced in \eq{E_1-explicit-2}. We need to
evaluate $\langle \a,V_s\rangle$ for $s\in\ms$ up to second order in
the fluctuations about the critical point in $E_1$, which is
non-tangential to the gauge orbit $\cO_{\rm max}(\vec n)$. For this, we
introduce real linear coordinates $x^i, y^i$, $i=1,\dots,d_1$ on $E_1$
such that a generic vector $V_\Psi \in E_1$ is parametrized as
$V_\Psi = \big(x^i\, K_i\,,\, y^i\, \cJ(K_i)\big)$.
Then $\gamma^i = \ddd x^i$ and $\gamma^{i+d_1}=\ddd y^i$ for
$i=1,\dots,d_1$. We can choose coordinates on $T_C\cO$ such that
$G_{ij} =2\,\Tr\big(X_i\, X_j^\dagger\big)$ is diagonal. One then finds 
\be
\langle \a,V_s\rangle = -\Tr\big({\rm ad}_s(V_\Psi)~{\rm ad}_{C_0}(V_\Psi)
\big)= \big(x^i\,,\,y^i\big) ~\tilde M_{ij}(s) ~
\left(\begin{array}{c} x^j \\ y^j\end{array}\right)
\ee
to second order, where
\be
\tilde M_{ij}(s) ~=~
(s_k-s_l)\, c_{kl}\, \left(\begin{array}{ccc} G & 0  \\ 0 & G 
\end{array}\right)_{ij}
\label{tildeMinblock}\ee
is a symmetric matrix and
\be
c_{kl} =\frac N2 \,\frac{n_l-n_k}{n_k\,n_l} \ .
\label{ad-C0-explicit}
\ee
One finds similarly
\be
\pfaff (A)
= 2^{d_1}~\sqrt{\det\big(\tilde M(s)\big)}~
\prod_{k > l}\,(s_k-s_l)^{1-|n_k-n_l|} \ .
\label{pfaffAtildeM}\ee
These pfaffians  are the typical representatives of fluctuations in
equivariant localization~\cite{szaboloc}, as discussed in
Section~\ref{DHThm}. Using the analytic continuation $t_1 \to
t_1-\ii\varepsilon$ and a suitable orientation of $E_1$ as before, we
can now evaluate the oscillatory gaussian integral
\be
\int_{E_1}\,\prod_{i=1}^{d_1}\,\ddd x^i~\ddd y^i ~ t_1^{d_1} ~
\e^{-\ii t_1\,\langle \a,V_s\rangle}
= \left(\frac{\pi}{\ii}\right)^{d_1}\,\frac 1{\sqrt{\det
\big( \tilde M(s)\big)}} \ .
\label{gaussintE1}\ee

Finally, putting the results \eq{Zdeg-4}, \eq{locmhms},
\eq{pfaffAtildeM} and \eq{gaussintE1} together, we may evaluate the
large $t_1,t_2$ limit of the desired symplectic integral \eq{Zmaxt1t2}
to obtain
\bea
Z'_{\rm max} &=& \frac1{\vol(G)}\,\int_{\mg\times {\cal F}_{\rm max}}\,
\Big[\,\frac{\ddd\phi}{2\pi}\,\Big]~
\exp\Big(\ddd(t_1\, \a+t_2\,\a'\,) - \ii\langle t_1\,\a+t_2\,
\a',V_\phi\rangle\Big)\nn\\ && \qquad\qquad\times~
\e^{-\ii \Tr(C_0\, \phi) -\frac{g'}{2}\,\Tr(\phi^2)} \nn\\[4pt]
&=& \frac1{\vol(G)}\, \left(\frac{\pi}{\ii}\right)^{d_1}\,
 \frac{\ii^{d_0}}{(2\pi\, N^2)^{d_0}}\,
\int_{\ms}\,\Big[\,\frac{\ddd s}{2\pi}\,\Big]~ \prod_{k > l}\,
(s_k-s_l)^{|n_k-n_l|+1}~
\frac{\pfaff(A)}{\sqrt{\det\big(\tilde M(s)\big)}} \nn\\
&& \qquad\qquad\times~
\frac 1{N^{d-d_0-d_1}}\, \int_{\cO_{\rm max}(\vec n)} \,\Big(\,
\bigwedge_{j=1}^{d-d_0-d_1}\,\lambda^j\Big)~ \wedge ~\Big(\,
\bigwedge_{i=1}^{2d_0}\,\b^i \Big)~
\e^{-\ii \Tr(C_0\,s) -\frac{g'}{2}\, \Tr (s^2)}\nn\\[4pt]
&=&\frac1{\vol(G)}\,\frac{\ii^{d_0-d_1}}{(2\pi)^{d_0-d_1}}\,
\prod_{k=1}^n\, \sqrt{n_k}\, \int_{\R^n}\,\Big[\,\frac{\ddd s}{2\pi}\,
\Big]~\Delta(s)^2~
\e^{-\ii \Tr(C_0\, s) -\frac{g'}{2}\, \Tr (s^2)}\nn\\
&& \qquad\qquad\times~
\frac 1{N^{d+d_0-d_1}}\, \int_{\cO_{\rm max}(\vec n)} \,\Big(\,
\bigwedge_{j=1}^{d-d_0-d_1}\,\lambda^j\Big)~ \wedge ~\Big(\,
\bigwedge_{i=1}^{2d_0}\,\b^i \Big)
\label{Zmax-s1}\eea
where we have transformed the integration over $\phi =
s=\diag(s_1~\one_{n_1},\dots,s_n~\one_{n_n}) \in \ms$ to an integral
over $s=(s_1,\dots,s_n)\in\R^n$. We can carry out the integral over
the moduli space $\cO_{\rm max}(\vec n)$ by observing again
\be
\frac 1{N^{d+d_0-d_1}}\, \int_{\cO_{\rm max}(\vec n)}\,\Big(\,
\bigwedge_{j=1}^{d-d_0-d_1}\,\lambda^j\Big)~ \wedge ~\Big(\,
\bigwedge_{i=1}^{2d_0}\,\b^i \Big)
= \int_{G/S}\,\bigwedge_{j=1}^{d+d_0-d_1}\,\eta^j
= \frac{\vol(G)}{\vol(S)} \ ,
\label{Omaxint}\ee
where $\cJ^*(\lambda^i) = \eta^i$ are left-invariant one-forms on the
gauge group $G$. Note that \eq{Omaxint} includes the integral over $E_0$,
and $\dim_\R (\mg \ominus \ms) = d + d_0 - d_1$. 
We also have $\vol(S) = \prod_k\, 2\pi\, \sqrt{n_k}$ in our
metric on $\ms$, since $S=\prod_k\,U(1)\otimes\one_{n_k}$, 
and $C_0(n_i) = \frac {N}{2n_i}~\one_{n_i}$.
Using furthermore $d_0- d_1 = n^2-n$ which is an even integer,
we may then bring (\ref{Zmax-s1}) into the form
\be
Z'_{\rm max} = \frac{\ii^{n^2-n}}{(2\pi)^{n^2+n}}\,
\int_{\R^n}\, [\ddd s]~ \Delta(s)^{2}~
\e^{-\ii \Tr(C_0\,s) -\frac{g'}{2} \,\Tr (s^2)}
\ee
which immediately leads to (\ref{Zmax-s2}).

\subsection*{Acknowledgments}

We would like to thank the organizers of the Orsay meeting for the
invitation and hospitality in a stimulating atmosphere. The work of
H.S. is supported by the FWF project P18657. R.J.S.\ was supported in
part by the EU-RTN Network Grant MRTN-CT-2004-005104.

\end{document}